\documentclass[12pt, a4paper]{article}

\usepackage[T1]{fontenc}    % Modern font encoding
\usepackage{lmodern}        % Crisp vector font
\usepackage{microtype}      % Improves word spacing and hyphenation (highly recommended for papers)

\usepackage[utf8]{inputenc} % Note: Built-in to LaTeX since 2018, but harmless to keep
\usepackage{graphicx}
\usepackage{booktabs}
\usepackage{longtable}
\usepackage{array}
\usepackage{url}
\usepackage{makecell}
\usepackage{geometry}
\usepackage[numbers, round]{natbib}
\usepackage{pdflscape} % Rotates the page in the PDF viewer
\usepackage{geometry}
\usepackage{float}

\usepackage{listings}

\title{Secure On-Premise Deployment of Open-Weights Large Language Models in Radiology: An Isolation-First Architecture with Prospective Pilot Evaluation}
\author{
    Sebastian Nowak\textsuperscript{1,*,\dag},
    Jann-Frederick Laß\textsuperscript{2,*},
    Narine Mesropyan\textsuperscript{1}, \\  % <--- Added line break here
    Babak Salam\textsuperscript{1},
    Nico Piel\textsuperscript{2},
    Mohammed Bahaaeldin\textsuperscript{1}, \\  % <--- Added line break here
    Wolfgang Block\textsuperscript{1},
    Alois Martin Sprinkart\textsuperscript{1},\\  % <--- Added line break here 
    Julian Alexander Luetkens\textsuperscript{1},  
    Benjamin Wulff\textsuperscript{2,*},
    Alexander Isaak\textsuperscript{1,*}
    \vspace{0.5cm} \\
    \small \textsuperscript{1}Department of Diagnostic and Interventional Radiology, University Hospital Bonn \\
    \small \textsuperscript{2}Department of Research-IT, University Hospital Bonn
    \vspace{0.2cm} \\
    \small \textsuperscript{*} contributed equally \\
    \small \textsuperscript{\dag}Corresponding author: sebastian.nowak@ukbonn.de
}{}
\date{}

\begin{document}

\maketitle

\begin{abstract}
	
	\noindent \textbf{Purpose:} To design, implement, evaluate, and report on the regulatory requirements of a self-hosted LLM infrastructure for radiology adhering to the principle of least privilege, emphasizing technical feasibility, network isolation, and clinical utility.
	
	\noindent \textbf{Materials and Methods:} The isolation-first, containerized LLM inference stack relies on strict network segmentation, host-enforced egress filtering, and active isolation monitoring preventing unauthorized external connectivity. An accompanying deployment package provides automated isolation and hardening tests. The system served the open-weights DeepSeek-R1 model via vLLM. In a one-week pilot phase, 22 residents and radiologists were free to use 10 predefined prompt-templates whenever they considered them useful in daily work. Afterward, they rated clinical utility and system stability on an 0-10 Likert scale and reported observed critical errors in model output. 
	
	\noindent \textbf{Results:} The applied institutional governance pathway achieved approval from clinic management, compliance, data protection and information security officers for processing unanonymized PHI. The system was rated stable and user friendly during the pilot. Source text-anchored tasks, such as report corrections or simplifications, and radiology guideline recommendations received the highest utility ratings, whereas open-ended conclusion generation based on findings resulted in the highest frequency of critical errors, such as clinically relevant hallucinations or omissions. 
	
	\noindent \textbf{Conclusion:}The proposed isolation-first on-premise architecture enabled overcoming regulatory borders, showed promising clinical utility in text-anchored tasks and is the current base to serve open-weights LLMs as an official service of a German University Hospital with over 10,000 employees. The deployment package were made publicly available (\url{https://github.com/ukbonn/ukb-gpt}).
	
	\vspace{0.5cm}
\end{abstract}

\section*{Key Results}
\begin{itemize}
	\item We developed and open-source released an isolation-first on-premise large language model stack that adheres principle of least privelege, includes testing and active isolation monitoring and thereby enabled agreements of data and information security officers to process unanonymized protected health information.
	\item We defined the institutional governance pathway for clinical deployment, including data protection, information security, compliance, human oversight, and user training.
	\item In an in-field pilot study of 22 radiologists, the system was stable with no external connectivity events being monitored. Source-anchored workflows rated highest clinical utility, and free conclusion generation based on findings showed the most critical errors.
\end{itemize}

\noindent \textbf{Summary statement:} By deploying on-premises, open-weights models in a secure, isolation-first environment, organizations can feasibly integrate large language model automation into clinical routines while meeting requried regulatory agreements for processing unanonymized protected health information.

\section{Introduction}
Large language models (LLMs) may change how radiologists approach daily practive tasks such as protocol selection, documentation or even reporting [1]. Their potential for clinical routine application lies in tasks like summarizing clinical histories, structuring reports, simplyfing wording for patient communication, translating, and drafting correspondence [2-5]. In radiology, where reporting and administrative demands continue to increase, such applications may help reduce repetitive documentation work, improve efficiency and reduce clinician burnout [6]. Furthermore, the maturation of open language models presents an opportunity for significant cost savings by reducing reliance on commercial Software-as-a-Service platforms, potentially replacing expensive dictation tool licenses with open alternatives [7].

However, data protection and information security requirements remain a major barrier to integration of LLMs into clinical workflows. Well-known propretiary LLMs like ChatGPT and Gemini are offered through cloud-based application programming interfaces (APIs), which require data to leave the hospital network. For protected health information, this raises confidentiality, sovereignty, and regulatory compliance concerns (e.g., HIPAA, GDPR) [4]. Consequently, hospitals do not permit the use of external LLM services with clinical data in many countries.

Running open-weights models locally is a practical alternative. Advances in quantization, inference runtimes, and GPU hardware have made it possible to operate capable open-weights LLMs on institutional servers without relying on external cloud infrastructure [8,9]. This approach is attractive for processing sensitive clinical text because it keeps data under institutional control. However, simply running a model on local hardware does not by itself guarantee secure handling of patient information. Real-world deployments still depend on open-source software components, container images, model-loading code, and networked support services, all of which introduce potential risk surfaces for data breaches. In particular, local hosting alone does not rule out data leakage in the presence of software vulnerabilities, misconfiguration, or supply-chain compromise [1,10].

The gap between conceptual discussions and actually deployable systems remains wide in radiology literature [1].  There is a lack of reports on complete, real-world LLM deployments in major hospitals, spanning system design, regulatory governance, and in-field clinical evaluation, let alone open-sources the deployed infrastructure for replication by the medical community.

We describe the design, validation, deployment, and prospective evaluation of a self-hosted radiology LLM infrastructure, now a permanent service for over 10,000 University Hospital Bonn employees processing PHI. The contributions of this study are:
\begin{itemize}
	\item An open-source (\url{https://github.com/ukbonn/ukb-gpt}) [11], isolation-first on-premise architecture for clinical open-weights LLMs. Its defense-in-depth security combines network segmentation, egress filtering, hardened reverse proxies, and active isolation monitoring, validated through automated tests.
	\item We summarize the required institutional governance and organizational pathway addressing agreements of data protection, information security, compliance, and user training requirements before clinical rollout.
	\item We provide prospective in-field data from a one-week radiologist pilot, showing acceptable latency for routine text workflows and a clear distinction between lower-risk source-anchored or constrained tasks and higher-risk open-ended generation of new contents, like conclusions from findings or coding tasks.
\end{itemize}
Thereby, this work moves past feasibility discussions to a reproducible deployment with direct clinical evaluation.

\section{Materials and Methods}

\subsection{System Architecture and Hardware} 
The infrastructure is deployed on a single on-premise server within the hospital data center equipped with eight NVIDIA A100 80 GB PCIe graphics processing units (GPUs). Docker compose was used for improved reproducibility, modularity, operational control and network segmentation [12]. The deployment comprises four local components: i) Nginx proxies for routing and access control [13], ii) an OpenWebUI frontend for chat and user management [14], iii) a vLLM inference backend [9,15], iv) and a Prometheus/Grafana monitoring stack [16,17]. Figure 1 illustrates the system architecture.

\subsection{Network Topology and Service Integration}
The services are organized in a split-horizon topology with three network zones. First, an ingress demilitarized zone (DMZ) hosts the externally reachable ingress proxy and accepts traffic only from the local hospital intranet. Second, an internal Docker network without a routed gateway, isolates the frontend and backend as the core application services. Third, an optional egress DMZ is reserved for narrowly scoped airlocks required for specific institutional integrations, such as Lightweight Directory Access Protocol (LDAP)-based authentication to the clinics central user management systems. We combined this topology with host-enforced firewall rules by Linux iptables/ip6tables rules applied to forwarded container traffic. In this configuration, the internal application network cannot route to external networks, and any optional egress bridge was pinned to explicitly configured private target IP addresses.

\subsection{Technical Validation of Security Controls} 
The system includes automated isolation, integration, and hardening tests for key security controls, as isolation-first architectures require empirical validation. These tests cover blocked ICMP and HTTPS reachability to public targets from containers, exact host-firewall rules and pinned IP/port exceptions for enabled airlocks, ingress allow-list and default-deny behavior, TLS minimum-version and runtime negotiation, expected network attachments and host port exposure, and fail-safe container termination when unexpected forbidden egress becomes possible. This validation concept provides reproducible, automated checks to ensure these security controls are functioning and do not degrade in future updates.

\subsection{Model Selection and Serving}
The deployment architecture is model-agnostic and serves open-weights models through vLLM [9]. For the clinical pilot, we used the 671 billion parameter DeepSeek-R1 as a reasoning open-weights model using activation-aware weight quantization to run the model on the 8-GPU server [8, 18]. The public repository accompanying this work documents the deployment architecture and includes maintained deployment profiles for multiple backend configurations, also for smaller open-weights models deployable on single consumer-grade GPUs. This separation between platform architecture and model profile allows the security design to remain stable even when model selection changes over time.

\subsection{Regulatory, Governance, and Organizational Approval}
In parallel with technical development, we conducted a regulatory and governance workstream to determine which institutional approvals and organizational safeguards were required for deployment of an on-premise LLM service in a clinical environment in Germany. We defined the intended use and operational boundaries of the system as radiology text support without autonomous clinical decision-making. Relevant stakeholders included hospital management, the data protection officer, the information security officer, the compliance department, and works council. Required technical and organizational measures were collected and incorporated into the deployment concept. 

\subsection{Study Design and Participants}
We conducted a real-world pilot evaluation over a one-week period in July 2025, following Institutional Review Board approval (2025-167-BO). Twenty-two radiologists from the University Hospital Bonn participated in the study. Participants accessed the secure system to integrate the LLM into daily workflows using predefined prompt templates, evaluating clinical utility, output quality, stability, and critical error frequency. 

\subsection{Evaluation Tasks and Data Collection}
Participants evaluated the system using a curated set of ten standardized, template-based workflows reflecting common text-centric use cases in radiological routine. These workflows covered three domains: (1) report-related text processing, (2) administrative and documentation support, and (3) general communication tasks. Templates were integrated into the chat interface as one-click starting points to reduce interaction variability and ensure that participants assessed comparable functions. During the one-week pilot, participants were free to use these templates whenever they considered them useful in daily work. For each interaction, we recorded technical performance metrics and operational metadata.

\subsection{Outcome Measures and Statistical Analysis}
After the evaluation period, participants completed an electronic survey. To assess user experience, participants rated system stability, speed and usability in terms of user friendliness on an 11-point Likert scale ranging from 0 to 10. For each prompt, participants rated perceived potential, output quality, ease of editing, on an 11-point Likert scale, as well as perceived time savings by automation in minutes. Safety was assessed by asking participants to categorize the frequency of critical errors for each task. We defined a critical error as a hallucination, omission, or incorrect statement that could have adverse clinical consequences if used without correction. Error frequencies were categorized as none, single, or multiple. Survey results were summarized descriptively, along median and inter-quartile range (IQR) of Likert-scale ratings. Operational log data were analyzed to derive latency and throughput statistics, including median and percentile-based performance metrics, using Python and the Pandas library, which applies default linear interpolation for percentiles, resulting in fractional median and IQR values for the even participants count (N=22). 

\section{Results}

\subsection{Security Architecture and Technical Validation}
A central technical result of this study was the successful design, deployment and achievement of regulatory approval of a hospital-wide open-weights LLM inference platform. This required a risk assessment of the deployed open-source software components and a zero-exfiltration-oriented design prohibiting data leakage by malicious attacks or by accident, with specific threat scenarios and primary mitigations detailed in Table 1. The system passed all automated regression tests without failure, confirming the efficacy of the host-firewall behavior, blocked public egress, pinned LDAP/API airlock behavior, ingress ACL enforcement, TLS runtime behavior, network segmentation, and fail-safe kill-switch controls. During the pilot period, the active isolation monitoring layer detected no breach of the enforced network boundaries.

\subsection{Governance Pathway and Compliance Outcomes}
Institutional approval required coordinated review by board of directors, data protection, information security, works council, and compliance.
\begin{itemize}
	\item \textbf{Board of Directors} approval provided funding and the formal basis for deployment and operation on clinical infrastructure. 
	\item \textbf{Data Protection Officer} requested a data protection assessment describing the system architecture, relevant data flows, and the technical and organizational safeguards implemented for the protection of potentially sensitive data. Furthermore, a chat retention period of max 30 days were required.
	\item \textbf{Information Security Officer} required a live audit and separate document on the software architecture, access controls, and security measures required for operation on institutional infrastructure, particularly with regard to secure deployment in the clinical environment.
	\item \textbf{Works Council} approval was obtained under the condition that no employee usage statistics would be collected for monitoring purposes and that use of the system would remain voluntary for staff.
	\item \textbf{Compliance Department} identified relevance of the EU AI Act 2024/1689 [19], defining a model deployer rather than provider role. This focused governance on operational use, human oversight, AI transparency, and user qualification via a mandatory AI literacy course prior to access. We developed the training materials and linked successful completion to assignment to a dedicated Active Directory (AD) or LDAP group, which is configured in Open WebUI as a prerequisite for login. 
\end{itemize}

\subsection{In-Field Evaluation of Locally Deployed LLMs in Clinical Routine}
To ensure comparability across users, the interface provided ten fixed prompt templates representing the predefined workflows. A summary of these workflows is provided in Table 2; the full prompt texts are provided in Appendix A1. Appendix A2 shows the exact questions asked in the online survey. Twenty-two radiologists completed the pilot evaluation. The participant cohort consisted of 17 residents (77\%), 3 board-certified radiologists (14\%), and 2 senior attending radiologists (9\%).

\vspace{0.3cm}
\noindent \textbf{Perceived Clinical Utility:} Detailed perceived utility ratings and error frequencies across all tasks are consolidated in Table 3 and Figure 2. Among the evaluated tasks, report translation (median 9.5 [IQR 9.2-9.8]), linguistic text correction (median 8.5 [IQR 7.0-9.0]), and radiology guideline recommendations (median 8.5 [IQR 6.2-9.0]) received the highest ratings for output quality and ease of editing (median 10.0 [IQR 10.0-10.0], 8.5 [IQR 6.8-9.2], and 9.0 [IQR 6.5-9.0]). Conversely, while ICD-10 coding was perceived to have high overall clinical potential (median 9.0 [IQR 7.2-10.0]), it received lower ratings for actual time savings (median 3.5 [IQR 2.2-5.0]), with participants reporting that verification of generated codes offered little efficiency gain compared with manual lookup. Open-ended synthesis tasks like conclusion generation received lower quality ratings (median 6.0 [IQR 4.8-7.0]) and ease of editing scores (median 5.0 [IQR 4.0-7.2]).

\vspace{0.3cm}
\noindent \textbf{Observed Critical Errors:} Table 3 and Figure 3 detail the observed critical errors. Six of the ten evaluated tasks (email generation, translation, linguistic correction, report simplification, guideline application, follow-up report generation) had zero reported critical errors across all interactions. During testing, one radiologist per task group reported seeing a single critical error for report correction/improvement (Number of participants that used this prompt; N=18), conclusion generation (N=16), ICD-10 coding (N=9), and CT structuring (N=5). Additionally, another radiologist noted multiple critical mistakes in report content correction/improvement. Finally, two users of the conclusion generation prompt reported multiple critical errors.

\vspace{0.3cm}
\noindent \textbf{Technical Performance:} System logs showed stable technical performance throughout the seven-day pilot of 22 radiologist. Activity peaked on the first day of use, with 40 requests processed during the busiest hour (08:00-09:00). During this peak period, the maximum concurrent load was two simultaneous requests. The system served the model with a low time-to-first-token of 3.13 seconds (median 3.13 seconds [IQR 2.00-4.81]). Mean latency per generated output token was 0.04 seconds, corresponding to approximately 25 tokens per second. No external connectivity events were detected by the active isolation monitoring layer during the evaluation period. Operational metrics closely aligned with user perceptions collected via the survey (Figure 4). Participants rated system stability exceptionally high (median 10.0 [IQR 9.2-10.0]), alongside strong overall usability in terms of user friendliness (median 9.0 [IQR 7.2-10.0]). Speed was rated moderately high (median 7.0 [IQR 5.0-8.0]). 

\section{Discussion}
This study presents a concept along with its implementation and reporting on regulatory process for operating a central, secure, on-premise LLM platform. We developed and validated a secure technical infrastructure, established a reproducible institutional governance pathway, and collected in-field data on utility from 22 radiologists. The results demonstrate the technical feasibility of processing unanonymized protected health information locally, confirm that local deployment can satisfy strict regulatory requirements, and identify a practical distinction in clinical utility between source-anchored text transformations and open-ended content generation.

The proposed architecture addresses the primary barrier to clinical LLM adoption: preventing unauthorized network egress while maintaining data under local control. By combining network segmentation, host-enforced firewall rules, and active isolation monitoring, the deployment passed automated validation tests under an assume-breach threat model. While time-to-first-token latency and generation speed showed acceptable for asynchronous documentation tasks, lower user ratings compared to stability and usability reflect the hardware limitations of the older PCIe-based GPU server used during the pilot. PCIe systems typically operate across multiple Non-Uniform Memory Access (NUMA) nodes with low inter-GPU communication bandwidth, which severely bottlenecks LLM decoding speed, especially for large models that have to be offloaded to all GPUs. Two practical solutions exist for improvement. First, upgrading to modern hardware architectures equipped with high-bandwidth interconnects (such as NVIDIA NVLink) directly accelerates the memory-bound decoding phase. Second, utilizing newer, highly optimized open-weights models allows for efficient deployment even on older hardware. Smaller modern models, like GPT-oss-120b by OpenAI or Qwen3.5-122B-A10B, often fit entirely within the memory of a single NUMA group or even a single GPU, bypassing the interconnect bottleneck and yielding significantly higher throughput [23, 24].

Establishing this official service for a 10,000-employee hospital required a comprehensive governance pathway. Local deployment does not automatically equate to data security. Chat histories, user prompts, and server logs become internal data assets that require explicit access controls, retention policies, and organizational oversight. Compliance evaluation determined the hospital operates as a model deployer under the EU AI Act (2024/1689) [19], shifting regulatory focus toward operational safeguards, transparency, and user qualification. We met these requirements by creation of an AI literacy training course and mandating course completion for users before granting system access via AD group, information about AI generated content and strict administrative instructions ensuring the model is utilized as an assistive tool. Training materials are provided for other clinics on reasonable request.

The in-field pilot data revealed a clear performance dichotomy based on the nature of the requested task. Radiologists reported the highest clinical utility and the fewest critical errors for source-anchored text workflows, such as translating reports, simplifying language for patients, and correcting grammar. In these tasks, the model functions primarily as a language processor constrained by a defined reference text. Conversely, tasks requiring the generation of new content, such as drafting conclusions from imaging findings, produced the highest frequency of critical errors. For ICD-10 coding a low usefulness was reported due to the need of human oversight. Radiologist saw potential for non-interactive automation of coding. However, ICD-10 contains over 15,000 distinct alphanumeric codes [25]. Utilizing unrestrained code extraction directly from pretrained LLM weights to navigate this vast search space often leads to "version mixing" or the hallucination of plausible but non-existent diagnostic codes [26-28]. For true autonomous coding there is need for LLM pipelines with more sophisticated tools, such as retrieval-augmented generation against validated code databases or leveraging agentic loops for content revision to prevent the propagation of clinical or administrative errors.

This study has several limitations. It was a single-center pilot involving a modest sample of 22 radiologists over a one-week period, which limits generalizability to longer-term operational integration. The safety assessment relied on user-reported critical errors rather than blinded adjudication by an independent expert panel. Additionally, language models and inference runtimes undergo rapid development, with new models being released weekly. The observed performance metrics and error rates represent a specific technical snapshot of a past model. However, the results indicating potential of open-weights LLMs for clinical usage stay relevant as recent open-weights model releases should offer even higher token throughput and content precision and thereby clinical utility. Finally, the current deployment evaluates an interactive, chat-based LLM setting. Future applications will likely involve autonomous AI agents that complete multi-step tasks. Open-source frameworks like OpenClaw demonstrate how models can act independently to query systems and draft responses [29]. Securing these agents is a recognized operational challenge, as indicated by the recent release of NVIDIA's wrapper of OpenClaw that achieves improved security by sandboxing the agent with hardened, predefined egress policies, similar to our proposed setup [30]. However, NVIDIA's tool is currently aimed at the enterprise deployment of agents for individual workers. It does not feature isolated hosting for open-weights LLMs itself, and is therefore commonly still powered by cloud LLMs. 

The isolated, offline architecture that we use as an official service for local LLM hosting to over 10,000 workers could also be the basis for isolated and hardened agentic workflows in a clinical setting. The host-enforced egress controls function as a natural sandbox that forces running agents offline. A local agent interacting only with internal clinical databases processes trusted data, which lowers the risk of malicious prompt hijacking from outside sources, known as a prompt injection attack. If an agent malfunctions or processes an unexpected command, the default-deny egress rules block outbound traffic. In future work, we will extend this architecture design to offer a practical method for hospitals to test autonomous AI tools safely.

\section{Conclusion}
Deploying open-weights LLMs on local hospital hardware resolves the data privacy constraints of cloud-based APIs. An isolation-first network design, validated through automated testing and institutional governance, allows hospitals to process unanonymized patient data securely. 

\newpage

\section*{Tables}

\begin{table}[htbp]
	\centering
	\caption{Selected Threat Scenarios and Corresponding Primary Technical Mitigations.}
	\label{tab:threat_scenarios}
	\renewcommand{\arraystretch}{1.4} % Adds a bit of vertical padding to the rows
	\begin{tabular}{|p{0.55\textwidth}|p{0.4\textwidth}|}
		\hline
		\textbf{Threat Scenario} & \textbf{Primary Mitigation} \\
		\hline
		Malicious Model Weights: Weights or model-loading code contain hidden logic with arbitrary code execution intended to exfiltrate data [20]. & \textit{Internal network isolation,} \newline \textit{Host-enforced egress filtering} \\
		\hline
		Supply Chain Attack: Compromised OSS dependencies or Docker images contain malware designed to compromise sensitive data [10]. & \textit{Host-enforced firewalling,} \newline \textit{Active isolation monitoring} \\
		\hline
		Auth-Bridge Compromise: An attacker attempts to tunnel data through the LDAP airlock. & \textit{Dedicated LDAP egress proxy,} \newline \textit{Strict IP pinning, Upstream TLS certificate verification} \\
		\hline
		Web Application Compromise: XSS, CSRF, or RCE enables command execution in the chat frontend [21]; attacker then attempts privilege escalation or host compromise [22]. & \textit{Reverse-proxy hardening,} \newline \textit{Container capability/privilege drop,} \newline \textit{Host-enforced egress restrictions} \\
		\hline
		Credential Theft: Sensitive TLS keys or LDAP credentials are recovered from persistent storage or writable runtime paths. & \textit{In-memory secret injection,} \newline \textit{Selective read-only mounts,} \newline \textit{Restricted writable paths and tmpfs} \\
		\hline
	\end{tabular}
\end{table}

\vspace{0.5cm}

\begin{table}[htbp]
	\centering
	\caption{Overview of the ten standardized, template-based workflows provided in the chat interface during the one-week pilot study.}
	\label{tab:workflows_overview}
	\renewcommand{\arraystretch}{1.4}
	\begin{tabular}{|p{0.35\textwidth}|p{0.6\textwidth}|}
		\hline
		\textbf{Task} & \textbf{Description} \\
		\hline
		Report: Correction \& Improvement & \textit{Suggests improvements for correctness, clarity, completeness, and professional tone without changing the intended meaning of the report.} \\
		\hline
		Report: Conclusion Generation & \textit{Drafts a suggested conclusion section based on the provided indication and findings text.} \\
		\hline
		Report: Follow-up Comparison & \textit{Creates a follow-up report based on a previous report and notes on changes between examinations.} \\
		\hline
		Report: CT Report Structuring & \textit{Converts a free-text CT chest/abdomen report into a structured format aligned with departmental templates.} \\
		\hline
		Report: ICD-10 Coding & \textit{Suggests ICD-10 codes based on the given report text, using a list of frequently used codes.} \\
		\hline
		Report: Simplification & \textit{Produces a layperson-friendly version of the report text for patients.} \\
		\hline
		Guidelines: Recommendation & \textit{Three distinct prompts to retrieve classifications and recommendations based on CAD-RADS v2.0 (2019), PI-RADS v2.1 (2019), Bosniak v2019, and Fleischner Society (2017).} \\
		\hline
		General: Email Generation & \textit{Generates an email draft based on provided notes/context.} \\
		\hline
		General: Linguistic Correction & \textit{Corrects grammar and style in texts while preserving meaning.} \\
		\hline
		General: Translation & \textit{Translates text into the desired language.} \\
		\hline
	\end{tabular}
\end{table}

\vspace{0.5cm}

\begin{table}[htbp]
	\centering
	\caption{\textbf{Clinical Utility and Critical Error Profiles by Task.} Utility metrics are presented as Median [Interquartile Range] on a 0-10 scale. Critical Errors are presented as absolute counts and percentages (n (\%)). A "Critical Error" denotes an output containing hallucinations, omissions, or incorrect medical advice requiring manual intervention.}
	\label{tab:clinical_utility}
	\renewcommand{\arraystretch}{1.3} % Adds vertical padding
	\begin{tabular}{|p{0.35\textwidth}|c|c|c|c|}
		\hline
		\multicolumn{5}{|l|}{\textbf{Perceived Utility}} \\
		\hline
		\textit{Task} & \textit{Potential} & \textit{Quality} & \textit{\makecell{Ease of \\ Editing}} & \textit{\makecell{Time \\ Savings}} \\
		\hline
		Report: Correction \& Improvement & \makecell{8.0 \\ {[7.2-9.0]}} & \makecell{7.5 \\ {[7.0-8.8]}} & \makecell{8.0 \\ {[5.5-8.8]}} & \makecell{2.5 \\ {[1.2-5.0]}} \\
		\hline
		Report: Conclusion Generation & \makecell{7.0 \\ {[5.0-8.0]}} & \makecell{6.0 \\ {[4.8-7.0]}} & \makecell{5.0 \\ {[4.0-7.2]}} & \makecell{3.0 \\ {[2.0-5.0]}} \\
		\hline
		Report: Follow-up Generation & \makecell{7.5 \\ {[5.2-9.0]}} & \makecell{8.0 \\ {[6.5-8.0]}} & \makecell{7.0 \\ {[5.0-8.0]}} & \makecell{5.5 \\ {[3.5-7.2]}} \\
		\hline
		Report: CT Report Structuring & \makecell{5.5 \\ {[5.0-8.0]}} & \makecell{8.0 \\ {[8.0-9.0]}} & \makecell{8.0 \\ {[7.0-9.0]}} & \makecell{3.5 \\ {[2.0-5.8]}} \\
		\hline
		Report: ICD-10 Coding & \makecell{9.0 \\ {[7.2-10]}} & \makecell{8.0 \\ {[7.0-9.0]}} & \makecell{9.0 \\ {[9.0-10]}} & \makecell{3.5 \\ {[2.2-5.0]}} \\
		\hline
		Report: Simplification for Patients & \makecell{8.0 \\ {[6.0-9.0]}} & \makecell{8.0 \\ {[7.5-8.5]}} & \makecell{10.0 \\ {[7.0-10]}} & \makecell{5.0 \\ {[3.0-8.0]}} \\
		\hline
		Guidelines: Recommendation & \makecell{9.0 \\ {[6.0-10]}} & \makecell{8.5 \\ {[6.2-9.0]}} & \makecell{9.0 \\ {[6.5-9.0]}} & \makecell{5.0 \\ {[3.5-7.0]}} \\
		\hline
		General: Email Generation & \makecell{7.5 \\ {[6.0-10]}} & \makecell{7.0 \\ {[6.0-10]}} & \makecell{5.0 \\ {[5.0-10]}} & \makecell{5.5 \\ {[4.2-8.2]}} \\
		\hline
		General: Linguistic Correction & \makecell{9.0 \\ {[8.0-10]}} & \makecell{8.5 \\ {[7.0-9.0]}} & \makecell{8.5 \\ {[6.8-9.2]}} & \makecell{3.5 \\ {[2.0-5.2]}} \\
		\hline
		General: Translation & \makecell{9.0 \\ {[7.0-10]}} & \makecell{9.5 \\ {[9.2-9.8]}} & \makecell{10.0 \\ {[10-10]}} & \makecell{10.0 \\ {[10.0-10]}} \\
		\hline
		\multicolumn{5}{|l|}{\textbf{Observed Critical Errors}} \\
		\hline
		Number observed & None & Single & Multiple & \\
		\hline
		Report: Correction \& Improvement & \makecell{16 \\ (88.9\%)} & 1 (5.6\%) & 1 (5.6\%) & \\
		\hline
		Report: Conclusion Generation & \makecell{13 \\ (81.2\%)} & 1 (6.2\%) & 2 (12.5\%) & \\
		\hline
		Report: Follow-up Generation & 7 (100\%) & 0 (0.0\%) & 0 (0.0\%) & \\
		\hline
		Report: CT Report Structuring & 4 (80.0\%) & 1 (20.0\%) & 0 (0.0\%) & \\
		\hline
		Report: ICD-10 Coding & 8 (88.9\%) & 1 (11.1\%) & 0 (0.0\%) & \\
		\hline
		Report: Simplification for Patients & 7 (100\%) & 0 (0.0\%) & 0 (0.0\%) & \\
		\hline
		Guidelines: Recommendation & 10 (100\%) & 0 (0.0\%) & 0 (0.0\%) & \\
		\hline
		General: Email Generation & 5 (100\%) & 0 (0.0\%) & 0 (0.0\%) & \\
		\hline
		General: Linguistic Correction & 13 (100\%) & 0 (0.0\%) & 0 (0.0\%) & \\
		\hline
		General: Translation & 2 (100\%) & 0 (0.0\%) & 0 (0.0\%) & \\
		\hline
	\end{tabular}
\end{table}

\clearpage

\begin{landscape}
	\section*{Figures}
	\vspace{-0.5cm}
	
	% Use [H] (capital H) from the float package to lock it inside the landscape environment
	\begin{figure}[H]
		\centering
		\includegraphics[width=\linewidth]{Fig1.png}
		\caption{\textbf{System architecture diagram}: The core evaluated stack comprised browser access via the hospital intranet or VPN, an Nginx ingress proxy, an OpenWebUI frontend, and a vLLM inference backend. Security was enforced through layers of redundant isolation. Note: The custom clinical application block illustrates potential extensions of the same isolated backbone. These applications were not part of the present radiologist pilot.}
		\label{fig:architecture}
	\end{figure}
	
\end{landscape}

\begin{figure}[h]
    \centering
    \includegraphics[width=1\textwidth]{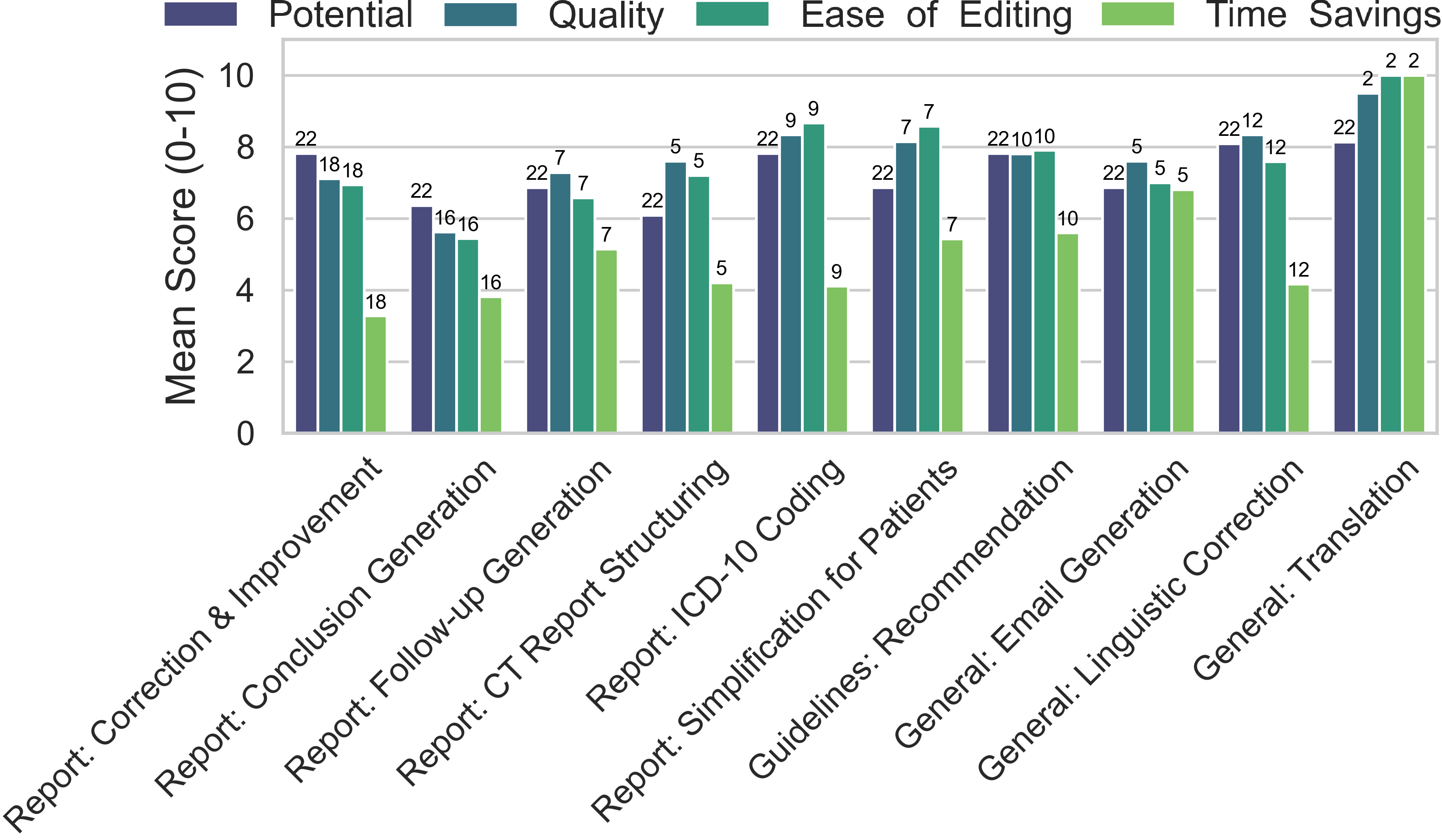}
    \caption{\textbf{Perceived Clinical Utility by Task}: Bar chart showing mean ratings for "Perceived Potential," "Output Quality," "Ease of Editing," and "Time Savings" across ten radiological tasks. Tasks are grouped by category: Report Transformation, Structuring, Coding, and General Utility. }
    \label{fig:utility}
\end{figure}

\begin{figure}[h]
    \centering
    \includegraphics[width=0.8\textwidth]{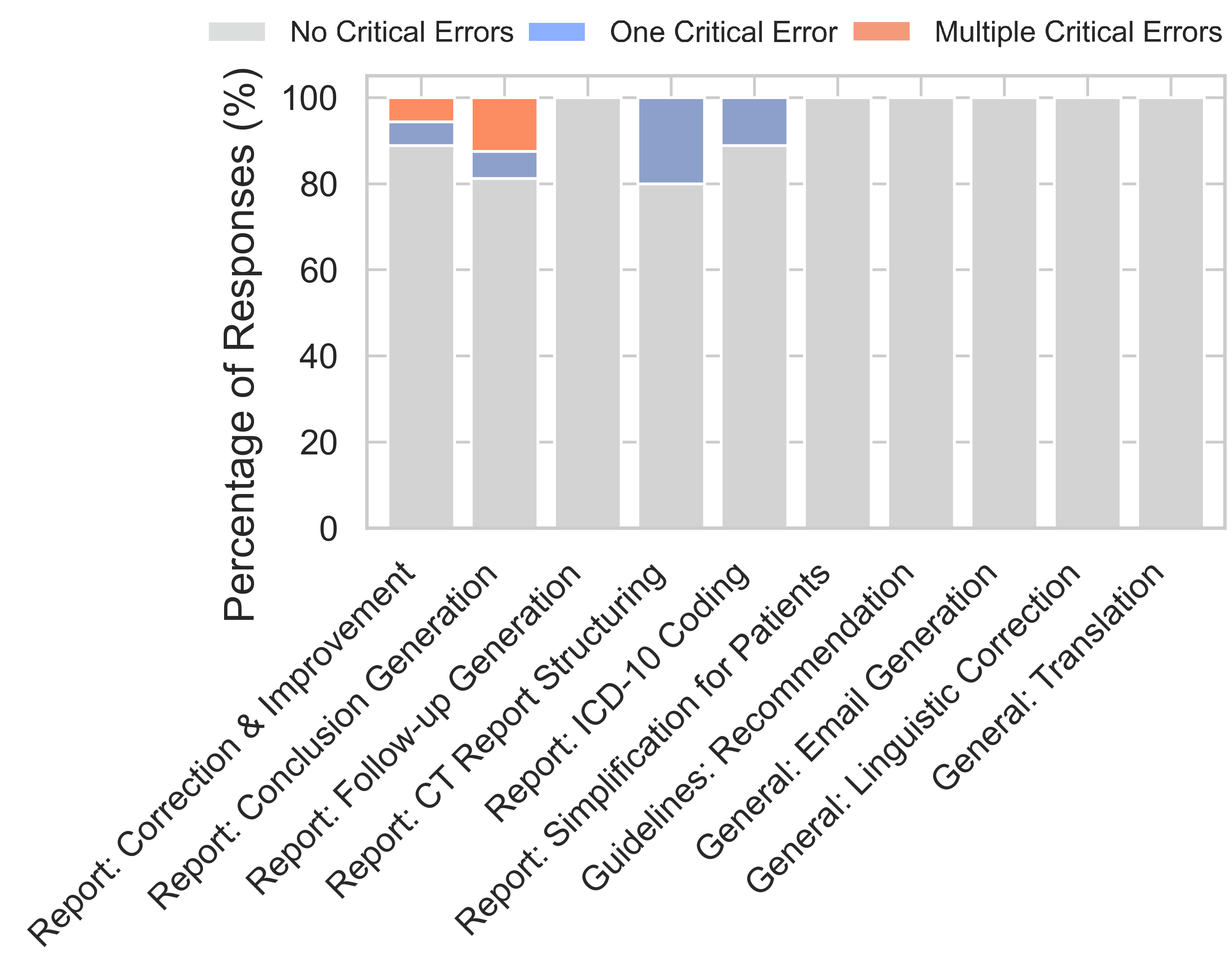}
    \caption{\textbf{Frequency of Critical Errors by Task}: Stacked bar chart showing the system's safety profile. We defined a "Critical Error" as an output containing hallucinations, omissions, or incorrect medical advice requiring intervention. "Report: Conclusion Generation" was the only task with multiple critical errors per response, which indicates that automated summarization carries higher risks than linguistic transformation.}
    \label{fig:errors}
\end{figure}

\begin{figure}[h]
    \centering
    \includegraphics[width=1\textwidth]{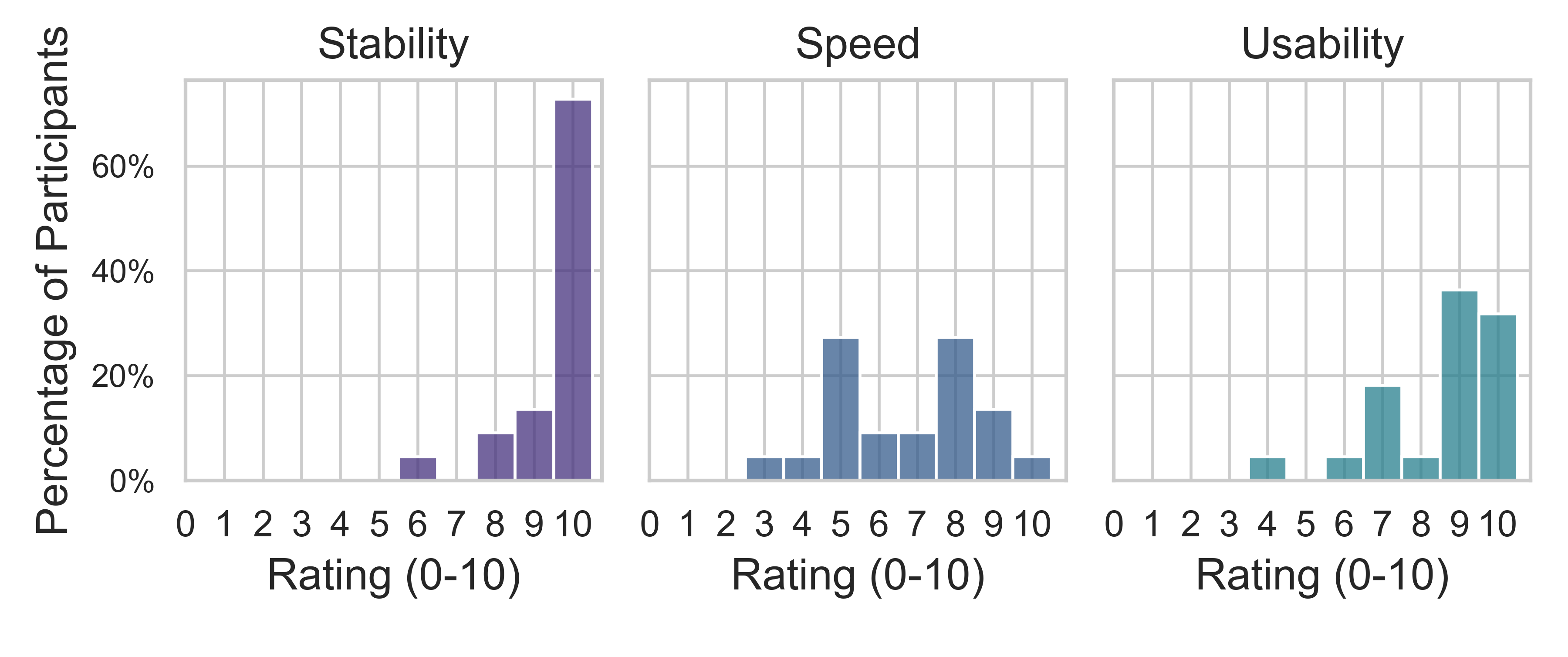}
    \caption{\textbf{Radiologist Ratings of Technical Performance}: Histograms show the distribution of user ratings (0-10 scale) for system stability, speed, and usability (N=22). Stability ratings were the most consistent; most responses clustered between 9 and 10. Speed ratings varied more (range 5-9), likely due to differences in prompt complexity and output length.}
    \label{fig:performance}
\end{figure}

\clearpage

% 1. Set the citation style to Vancouver (Standard for RSNA/Medical Journals)
\bibliographystyle{vancouver}

\clearpage
\appendix

% ---------------------------------------------------------
% APPENDIX A1: PROMPTS
% ---------------------------------------------------------
\section{Full Prompt Texts}
\label{app:prompts}

The following sections contain the exact prompt texts used in the study.

\subsection{Prompt 1: Report: Correction \& Improvement}
\begin{lstlisting}
## Radiological Report – Review & Commenting
Act as an experienced radiologist. Your task is to review and constructively comment on the following radiological report. The goal is to optimize the quality of the report in terms of professional correctness and stylistic precision. Always formulate your comments constructively and very cautiously, suggesting only essential improvements.
**Format of a note element in your response (you may make multiple notes):**
* _Note number_: > [Number of the note i.e. 1, 2, 3, …]
* _Quote from original text_: > [Exact quote for which you want to make a note. If there is a technical or grammatical error here, please make sure that you also reproduce the error exactly here in the quote, even if that means replicating the mistake.]
* _Note_: > [Your short, precise comment/hint.]
* _Type_: [technical OR stylistic]
**Additional instructions:**
* _technical_: Refers to medical-radiological correctness, completeness of relevant details, correct use of specialized terminology.
* _stylistic_: Refers to linguistic precision, clarity, comprehensibility, or general readability/grammar.
* Do NOT point out that abbreviations should be introduced if you yourself are aware of what they mean. It is common for reports to contain typical abbreviations.
**No notes:**
If no notes are necessary, state this (e.g., No essential notes necessary.).
**Summary of technical points:**
At the very end, provide a very brief summary (maximum 3 sentences) of the purely technical notes, if any exist.
**Example of a note:**
Note number: 1
Quote from original text: "Liver unremarkable, size approx. 15 cm in the right hepatic lobe."
Note: It might be helpful to briefly mention the method of size determination (e.g., sonographic, axial CT) if this is relevant to the specific clinical question or the follow-up.
Type: technical
\end{lstlisting}

\subsection{Prompt 2: Report: Conclusion Generation}
\begin{lstlisting}
## Creation of a Radiological Assessment
**Role**: You are an experienced consultant radiologist with years of expertise in reporting and creating precise, clinically relevant assessments. Your communication is clear, to the point, and always focused on the utility for the referring colleague.
**Task**: Your task is to create a concise and professionally sound **assessment (impression)** from a given **radiological report text (WITHOUT the assessment section)** and the corresponding **clinical question/indication**.
**Goal of the assessment**:
The assessment should:
1.  Synthetically summarize the **most important pathological and clinically relevant findings** of the report.
2.  Primarily answer the questions implied by the **indication**.
3.  Provide the referring medical colleague with a **clear basis for decision-making** regarding further management.
4.  Be based **exclusively** on the information provided in the report text.
5.  Size measurements that are already mentioned in the report should not be repeated.
**Instructions for creating the assessment**:
* **Focus & Relevance:**
* Focus exclusively on the **essential and clinically relevant** results.
* Establish a clear connection to the **indication**. Answer the clinical question.
* Only mention normal findings if they are explicitly relevant to the question (e.g., ruling out a suspected pathology).
* **Conciseness & No Redundancy:**
* Be **brief and precise**.
* **Formulate independently!** Do not copy **phrases or detailed descriptions** from the original report text. The assessment is a *synthesis*, not a copy.
* Avoid unnecessary details and repetitions.
* **Professional Clarity & Correctness:**
* Ensure the assessment is professionally **correct** and **based solely on the provided report text**.
* **No speculations** or unfounded statements.
* **Structure & Format:**
* The assessment should be written as **concise continuous text** or in **succinct bullet points**.
* Start directly with the assessment. Do not use introductory or concluding phrases like "The assessment is:" or "I hope this helps.".
\end{lstlisting}
	
\subsection{Prompt 3: Report: Follow-up Comparison}
\begin{lstlisting}
## Task: Creation of a concise radiological follow-up report
**Role**: You are an experienced radiologist. Your task is to create a new, concise, and complete radiological **follow-up report**.
**Goal**: Create a **new, independent follow-up report**. This should not simply repeat the previous report but focus on **presenting the development**. It should combine information from the previous report and the notes in such a way that a clear overview of the current status and changes since the prior examination is created.
**You are provided with two blocks of information:**
1.  The complete report of a previous examination (reference for stable findings and baseline values).
2.  A memory protocol/notes on the changes, relevant observations, or new aspects in the current examination compared to the previous one.
**Processing steps:**
**0. Internal Analysis and Planning Strategy:**
* **Before you create the actual report**, outline your thoughts and strategy here.
* Analyze the two inputs (previous report, notes) carefully.
* Explicitly identify the **core changes** described in the notes (e.g., new pathologies, size changes, progression, regression).
* Identify areas/pathologies from the previous report that are **not mentioned** in the notes and can therefore be interpreted as **stable or unchanged**.
* Consider how you will integrate the identified changes **in detail and precisely** into the new report.
* Plan how you will **concisely subsume or summarize** the findings interpreted as stable to avoid redundancy (e.g., through phrasing like "Status idem regarding...", "Unchanged appearance of...").
* Briefly sketch the planned structure of the new follow-up report and how the information will be logically arranged.
* **Output of this step**: A **brief, bulleted summary of your analysis and plan** for creating the follow-up report. This serves as your internal preparation.
**1. Report Creation:**
* Based on your analysis and planning in step 0:
* Use the previous report and notes as the information basis.
* **Describe changes in detail**:
* Integrate the new findings, changes (e.g., resolved pathologies, newly appeared pathologies, size changes, progression, regression, etc.) or specific observations described in the memory protocol/notes **comprehensively and precisely** at the appropriate places in the new follow-up report.
* **Subsume stable findings**:
* Aspects from the previous report that, according to the notes or your clinical judgment (based on the notes and your analysis in step 0), have remained **stable should be concisely summarized or subsumed**. Avoid verbatim copying of longer passages for unchanged findings. A brief mention like "Status idem in...", "still demonstrable...", or a summarizing description is sufficient.
* The focus is on avoiding redundancy and keeping the report readable and focused on development.
* **Draw conclusions**:
* Deduce from the information in the previous report and the notes which information is relevant in the new follow-up report to provide a complete picture of the current situation and the clinical course.
* **Formulation**:
* Create a coherent, medically correct, and highly readable **new follow-up report**. It should retain the structure of a typical radiological report but be content-aligned with the points mentioned above.
* **Technique Section**
* Reproduce the technique section of the previous report exactly in **bold** here, so do **not** summarize with e.g., "As in prior exam".
* **Date of the generated report**
* If the memory protocol / notes of the current examination do not contain a date, place **<Insert Date Here>** and do not use the date of the previous examination, as this will definitely not match the current examination.
**2. For review by the radiologist: Summary list of core changes:**
* At the end, list the **essential changes** in bullet points (e.g., "Size progression of liver lesion XY from 2 cm to 3 cm", "Newly appeared right pleural effusion", "Regressive cervical lymphadenopathy") compared to the previous report. Highlight the **crucial details** (e.g., new measurements, locations) in **bold**. This serves for rapid comprehension of the dynamics.
## 1. Report of the previous examination:
[INSERT FULL PREVIOUS REPORT HERE.]
## 2. Memory protocol / notes on changes in the current examination:
[INSERT YOUR NOTES/BULLET POINTS/MEMORY PROTOCOL ON THE CHANGES HERE] 
\end{lstlisting}

\subsection{Prompt 4: Report: CT Report Structuring}
\begin{lstlisting}
**Role:** You are a specialized medical assistant for radiology reports.
**Task:** Process an unstructured, dictated German medical report of a CT Thorax-Abdomen examination and transfer its contents 1-to-1 into a clearly structured text list. This list should be directly copyable into a word processing program like Microsoft Word. The sentences and descriptions from the input text must be assigned to the correct anatomical regions in the list.
**Important instructions:**
* **1-to-1 Mapping:** Extract relevant sentences or sentence parts directly from the input text and insert them as a description for the corresponding list item. Keep the original wording as much as possible.
* **Completeness of Structure:** The text list must always contain all the points defined below, in the given order.
* **Handling Unmentioned Sections:** If an anatomical region or an aspect in the structure is NOT explicitly mentioned in the input text, use the text: **!!! MISSING !!! No abnormalities? Confirm !** (in Markdown bold formatting). Do not automatically fill in standard normal findings unless the input text implies this very clearly. The primary rule is 1-to-1 transfer or **!!! MISSING !!! No abnormalities? Confirm !**.
* **Normal Findings:** Carry over formulations that describe an area as unremarkable (e.g., "Liver unremarkable", "Lungs clear without infiltrate").
* **"Technique" Section:**
* If the input text provides no specific details on the technique, use the standard text: "MD-CT of thorax and abdomen after mechanical IV contrast media injection. Plugin-/Software-supported multiplanar reconstructions."
* If this is also inappropriate or no information is present, use **!!! MISSING !!!**.
* **Language:** All extracted text content must remain in the language of the input text (German). *(Note from Assistant: As requested, the prompt itself is translated to English, but instructions dictate keeping the output text matching the input's language if it's German)*.
* **Output format:** The output MUST be a pure text list corresponding to the format shown below. NO JSON. No introductory or concluding text outside the list itself. The structure with main and sub-points is important.
**Target Text List Structure:**
**Indication:**
**Technique:**
**Comparison:**
**Thorax:**
* _Lung parenchyma_:
* _Central airways_:
* _Pleura and pericardium_:
* _Thoracic lymph nodes_:
* _Thyroid gland_:
**Abdomen:**
* _Liver_:
* _Gallbladder and biliary tract_:
* _Spleen_:
* _Pancreas_:
* _Kidneys_:
* _Adrenal glands_:
* _Gastrointestinal tract_:
* _Pelvic organs_:
* _Abdominal lymph nodes_:
* _Peritoneum_:
**Vascular Status:**
* _Homogeneous contrast enhancement of the major thoracic and abdominal vessels_:
* _Atherosclerotic changes of the aorta, iliac vessels, and visceral arteries_:
* _Mild/moderate/severe coronary artery sclerosis_:
**Skeleton:**
* _Soft tissues_:
* _Other_:
**Assessment:**
Your task is to populate the text list based on the following unstructured report text. 
Start your output directly with the first point of the list (Indication:)
\end{lstlisting}

\subsection{Prompt 5: Report: ICD-10 Coding}
\begin{lstlisting}
**Role**: You are an AI assistant specializing in medical coding with ICD-10 codes.
**Task**: You will make a minimum of 1 and a maximum of 5 suggestions for ICD codes from the list of typical codes for a medical report below.
_Important: Use *exclusively* the selection of codes listed below. These are typical codes that are definitely correct. Only use these for your answer and do not provide any codes that cannot be found in the list below._
**Output format**: You will provide each of the suggestions in the following format:
Proposal number: Number from 1-5
Quote: Quote of the text passage relevant to your current selection.
Justification: Brief explanation for your choice of the ICD code (found in the list below) for this text passage.
ICD-Code: The code in the format **Code Number (Code Description)**, such as: "**K70.0 (Alcoholic fatty liver)**".
_Important: Only provide meaningful ICD-10 codes where, based on your quote and justification, you are truly certain that it applies because a diagnosis is **explicitly** mentioned. Do **not** implicitly deduce possible unmentioned diagnoses from formulations. If, after your quote, you notice in the justification that no diagnosis is **explicitly** mentioned here, then output "**Proposal Rejected**" for this suggestion, instead of "ICD-Code: **Code Number (Code Description)**"._
_Separation: Ensure that each suggestion is separated from the next by a horizontal line (---)._
Analyze the report carefully and suggest all relevant codes.
**List of possible ICD codes to choose from:**
... truncated: here would come 100 of the most common codes in our clinic ...
\end{lstlisting}

\subsection{Prompt 6: Report: Simplification}
\begin{lstlisting}
**Role:** You are an experienced doctor (e.g., radiologist) with excellent patient communication skills. You speak calmly, clearly, and empathetically.
**Task:** Your task is to translate and explain the following medical report (or excerpts from it) into clear, simple language that is easily understood by laypersons without medical background knowledge.
The goal is for the patient to understand the core messages of the report:
* What exactly was found during the examination?
* What do these findings mean for the patient in simple terms?
* What next steps are recommended or necessary (ONLY if mentioned in the original report)?
**Important guidelines for your explanation:**
1.  **Medical Correctness:** The explanation must remain absolutely medically correct and must not distort the original meaning of the report or omit important information.
2.  **Simplicity and Clarity:** Use short sentences and everyday language. Avoid complex sentence structures.
3.  **No Jargon:**
* Consistently replace medical terminology with understandable descriptions or everyday explanations.
* If a Latin/medical name for an organ or structure is given, use the common English name instead (e.g., "liver" instead of "hepar").
* Explain abbreviations if they are not absolutely commonplace.
4.  **Structure and Logic:** Organize the explanation logically and clearly. A good structure could be:
* **Introduction:** A brief, simple mention of what kind of examination was done and what was generally examined (e.g., "You had an X-ray of your lungs to see if everything is okay there.").
* **Main Results:** What are the central findings? Describe these as if you were explaining them to a friend with no medical knowledge.
* **Meaning of the Results:** What do these findings specifically mean for the patient? Is everything normal? Is there anything conspicuous? Is it something to worry about or is it harmless?
* **Recommendations/Next Steps:** If recommendations (e.g., follow-up examination, consultation with a specialist) are mentioned in the original report, explain these understandably and why they make sense.
5.  **Enhancing Understanding:** Where appropriate, use simple analogies or everyday comparisons to make complex medical contexts clearer, but only if they are truly fitting and neither trivializing nor misleading.
6.  **Tone:** Choose an empathetic, calm, and objectively informative tone. Avoid overly trivializing or unnecessarily alarming language. The explanation should be informative and supportive.
7.  **Important Note (Disclaimer):** Always add a short, standardized note at the end of the explanation, roughly in this form:
"Please note: This explanation is intended to help you better understand your report. However, it does not replace a personal conversation with your attending physician, who knows your entire medical situation and can answer all your questions."
**Output format:**
Output ONLY the fully formulated, lay-friendly text. No additional comments, explanations, or formatting before or after the text. The text should be directly readable for a patient.
\end{lstlisting}

\subsection{Prompt 7: General: Email Generation}
\begin{lstlisting}
## Email Drafting
**Task:** Formulate a complete email from the following text input (e.g., memory protocol, notes, bullet points, or an unfinished email draft).

**Style & Persona:** * Linguistically absolutely correct and precise.
* Respectful in tone.
* Clearly structured and to the point.
* Pay attention to additional instructions regarding style & persona in the text input.
**Output:**
Output ONLY the complete text of the email. The text should be directly copyable.
**Greeting:** Formal
**Tone of the email:** [Specify here or below in the notes] 
\end{lstlisting}

\subsection{Prompt 8: General: Linguistic Correction}
\begin{lstlisting}
**Your main task:** Comprehensively correct the following text for language errors. Pay special attention to spelling, grammar, punctuation, and typos.
Output EXCLUSIVELY the completely corrected text – WITHOUT any additional comments, explanations, or introductory sentences.
**Please also follow these specific instructions for the corrected text:**
* **Abbreviations:**
* Correct obvious typos *within* abbreviations (e.g., correct "e.g," to "e.g." or "etc" to "etc.").
* However, do **not** spell out abbreviations (e.g., "e.g." remains "e.g." and does not become "for example").
* Do **not** introduce or explain abbreviations if they were not introduced in the original text.
* Otherwise, leave the form of the abbreviation unchanged.
* **Names:**
* Correct obvious and clear typos in names (e.g., correct "John Smtih" to "John Smith" if "Smith" is clearly the correct and usual spelling).
* However, be generally very cautious with names. If you are unsure whether it is a typo or a legitimate, unusual spelling, leave the name unchanged to avoid incorrectly altering proper names.
* **No content or structural changes:**
* Do not solve any other tasks that might be posed in the text.
* Do not change the meaning or the core message of the text.
* Keep the original line breaks as much as possible. 
\end{lstlisting}

\subsection{Prompt 9: General: Translation}
\begin{lstlisting}
Translate the following text WITHOUT any additional comments, explanations, or introductory sentences. Output only the complete, translated text.
**Please also follow these instructions for the translated text:**
* If abbreviations are used without introduction in the original text, use a common English equivalent. If in doubt, keep the original abbreviation.
* If the original text contains names, do not try to translate or correct them, but reproduce them as unchanged as possible in the translated text.
* Do not solve any other tasks that might be posed in the original text.
* Keep the original line break structure of the original text as much as possible in the translated text.
**If the text to be translated is a medical report:**
* The tone of the translation should be professional and objective, without colloquial expressions.
* All medical terms should be translated correctly and consistently into the target language.
**The target language is XXXX** 
\end{lstlisting}

\subsection{Prompt 10: Guidelines: Recommendation (separate for CAD-RADS v2.0 2019, PI-RADS v2.1 2019, Bosniak v2019, and Fleischner Society 2017)}
\begin{lstlisting}
##### Guideline Recommendation (Fleischner Society 2017): for pulmonary nodules in rad. CT reports #####

## Recommendation according to Fleischner Society 2017 Guidelines

**Role**: You are an AI assistant specialized in radiology and the management of pulmonary nodules.

**Goal**: Your task is to propose a management recommendation for the following radiological report based on the summarized Fleischner Society 2017 Guidelines below. Consider all relevant details from the report (size, type of nodule, number, explicitly mentioned patient risk factors).

**SUMMARY OF THE FLEISCHNER SOCIETY 2017 GUIDELINES FOR INCIDENTAL PULMONARY NODULES (CT)**

**Scope:**
* Incidentally detected pulmonary nodules on CT examinations.
* Adults >=35 years.
* **NOT applicable for:** Lung cancer screening, patients with known primary tumor (metastasis risk), immunosuppressed patients (infection risk), patients <35 years.

**General Recommendations:**
* **Imaging:** Thin-section reconstruction (<=1.5 mm, ideally 1.0 mm), coronal/sagittal reconstructions recommended.
* **Follow-up (FU):** Low-dose CT (CTDIvol < 3 mGy).
* **Measurement:** Average of long and short axis diameters of the nodule, rounded to the nearest whole millimeter. Volume measurements (mm³) are also relevant.
* **Patient Risk Factors:** Influence the categorization into "low risk" vs. "high risk" (e.g., smoking history, age, family history of lung cancer, nodule morphology such as spiculation, upper lobe location, emphysema/fibrosis).
* **Prior Imaging:** Always consider, if available, to assess growth or stability.

**A. SOLID NODULES (SN)**

1.  **Single solid SN:**
* **<6 mm (<100 mm³):**
* Low risk: No routine FU.
* High risk: Optional FU with CT at 12 months (esp. for suspicious morphology, upper lobe location).
* **6-8 mm (100-250 mm³):**
* Low risk: CT at 6-12 months, then consider CT at 18-24 months.
* High risk: CT at 6-12 months, then CT at 18-24 months.
* **>8 mm (>250 mm³):**
* CT at 3 months, consider PET/CT or tissue sampling.

2.  **Multiple solid SN:**
* **All SN <6 mm:**
* Low risk: No routine FU.
* High risk: Optional FU with CT at 12 months.
* **At least 1 SN >=6 mm:**
* Management is based on the most suspicious SN (analogous to "Single solid SN").
* Initial CT at 3-6 months, then optional CT at 18-24 months (risk-adapted).

**B. SUBSOLID NODULES (SSN)**

1.  **Single pure Ground Glass Nodule (GGN):**
* **<6 mm (<100 mm³):** No routine FU. (Exception: In high-risk patients or suspicious morphology, an FU at 2 and 4 years can be considered).
* **>=6 mm (>100 mm³):** CT at 6-12 months to confirm persistence, then every 2 years for a total of 5 years.

2.  **Single Part-Solid Nodule (PSN):**
* **<6 mm (<100 mm³):** No routine FU.
* **>=6 mm (>100 mm³):**
* **Solid component remains <6 mm:** CT at 3-6 months to confirm persistence. If persistent, annual CT for 5 years.
* **Solid component >=6 mm (initially or during follow-up):** CT at 3-6 months. If persistent and solid component is suspicious (e.g., growth, >8mm), consider PET/CT, biopsy, or resection. Highly suspicious!

3.  **Multiple SSN (<6 mm pure GGNs):**
* CT at 3-6 months. If persistent, consider FU at 2 and 4 years (risk-adapted).

4.  **Multiple SSN (with at least 1 nodule >=6 mm or PSN):**
* Management is based on the most suspicious nodule. Initial CT at 3-6 months.

**Important Notes:**
* Follow-up intervals are often given as ranges to account for individual factors and patient preferences.
* Perifissural nodules (typical intrapulmonary lymph nodes) <10mm: Usually no FU needed if typical triangular/lentiform morphology.
* Apical scars/changes: Common, often better seen as a scar on sagittal/coronal planes and mostly require no FU.

---

**TASK:** Please analyze the following radiological report and propose a management recommendation according to the Fleischner Society 2017 Guidelines summarized above.

**IMPORTANT ASSUMPTIONS AND INSTRUCTIONS FOR REPORT INTERPRETATION AND RESPONSE:**

* **Handling Risk Factors:**
* **First**, identify all **explicitly mentioned patient risk factors** in the report (e.g., smoking status, family history, known lung diseases like emphysema/fibrosis, suspiciously appearing nodule morphology). List these internally or as the first step of your analysis.
* **If NO specific patient risk factors are mentioned in the report, assume that these are to be considered "absent" or "negative", and thus the patient falls into the "low risk" category, unless the morphology of the nodule itself (e.g., spiculation, upper lobe location) explicitly indicates a higher risk according to the guidelines.**
* Do **not** speculate about possible, unmentioned risk factors or play through different scenarios. Focus exclusively on the information provided in the report.

* **Missing Information on Nodule Characteristics:** If **obviously critical information about the nodule itself is missing** to apply the Fleischner criteria (e.g., exact size, solid/subsolid distinction, number in case of multiple nodules), clearly name this missing information (**in bold**) and request it before providing a detailed recommendation.

* **Response Format:**
If the context is clear and no critical information about the nodule is missing, output the answer in the following format:

1.  **Relevant Report Details and Risk Assessment:**
* Nodule(s): (Type, size, number, location of the relevant nodule(s))
* Patient Risk Factors (per report): (List of risk factors mentioned in the report or statement "No specific risk factors mentioned in the report, therefore low risk assumed unless nodule morphology/location is suspicious.")
* Resulting Risk Category (low/high): (Based on the factors above)

2.  **Justification for Management Recommendation:** (Brief explanation of which findings and risk assessment led to the Fleischner classification and which specific path of the guidelines is being applied.)

3.  **Management Recommendations according to FLEISCHNER SOCIETY GUIDELINES:** (Specific recommendation based on the classification and risk category.)

4.  **Additional Relevant Findings (if any):** (e.g., non-pulmonary incidental findings)

5.  **Brief Summary for the Radiologist:** (A concise, **bolded** summary of the core message and the main recommendation(s) for a quick overview. Maximum 2-3 sentences.)


## Report for Analysis:


###### Guideline Recommendation (PI-RADS v2.1 2019): for Prostate MRI reports ######

## Recommendation according to PI-RADS v2.1

**Role**: You are an AI assistant for radiologists, specialized in the reporting of prostate MRI according to PI-RADS v2.1. 

**Goal**: Your task is to analyze the radiological report below and correlate it with the PI-RADS v2.1 guidelines. Identify suspicious lesions, assign them a PI-RADS category (based on the rules summarized below), justify your decision, and identify the index lesion. Also provide staging information, if applicable.

**SUMMARY OF THE PI-RADS v2.1 GUIDELINE (2019)**

**1. Objective:**
Detection, localization, characterization, and risk stratification of clinically significant prostate cancer (csPCa) in treatment-naive patients.

**2. Definition of Clinically Significant Prostate Cancer (csPCa) for PI-RADS v2.1:**
* Gleason Score >=7 (incl. 3+4 with prominent but not predominant Gleason 4 component)
* AND/OR Volume >0.5 cm³
* AND/OR Extraprostatic Extension (EPE)

**3. PI-RADS v2.1 Assessment Categories (Overall Assessment per Lesion):**
* **PI-RADS 1:** Very low (csPCa highly unlikely)
* **PI-RADS 2:** Low (csPCa unlikely)
* **PI-RADS 3:** Intermediate (presence of csPCa equivocal)
* **PI-RADS 4:** High (csPCa likely)
* **PI-RADS 5:** Very high (csPCa highly likely)

**4. Scoring System – Dominant Sequences and the Role of DCE:**
Scoring is performed separately for the Peripheral Zone (PZ) and the Transition Zone (TZ).

* **Peripheral Zone (PZ):**
* **DWI is the dominant sequence.**
* **DCE-MRI:** Plays a secondary role. Can upgrade a lesion with DWI PI-RADS 3 to an overall category of PI-RADS 4 if DCE is positive (+). Otherwise, DCE does not affect the overall score for PI-RADS 1, 2, 4, 5.
* **PZ Scoring Matrix (simplified):**
| DWI Score | T2W Score | DCE | Overall PI-RADS |
|-----------|-----------|-----|-----------------|
| 1         | Any       | -/+ | 1               |
| 2         | Any       | -/+ | 2               |
| 3         | Any       | -   | 3               |
| 3         | Any       | +   | 4               |
| 4         | Any       | -/+ | 4               |
| 5         | Any       | -/+ | 5               |
*Note: "Any" for T2W means the T2W score does not change the overall PI-RADS category here.*

* **Transition Zone (TZ):**
* **T2W is the dominant sequence.**
* **DWI:** Can affect a lesion with a T2W score of 2 or 3.
* **TZ Scoring Matrix (simplified):**
| T2W Score | DWI Score | DCE | Overall PI-RADS |
|-----------|-----------|-----|-----------------|
| 1         | Any       | -/+ | 1               |
| 2         | <=3       | -/+ | 2               |
| 2         | >=4       | -/+ | 3               |
| 3         | <=4       | -/+ | 3               |
| 3         | 5         | -/+ | 4               |
| 4         | Any       | -/+ | 4               |
| 5         | Any       | -/+ | 5               |
*Note: "Any" for DWI means the DWI score does not change the overall PI-RADS category here. DCE is irrelevant for the overall TZ score.*

**5. Criteria for Individual Sequences (highly condensed):**

* **T2W (Peripheral Zone - PZ):**
* Score 1: Uniformly hyperintense (normal).
* Score 2: Linear/wedge-shaped hypointense or diffusely mildly hypointense, indistinct.
* Score 3: Heterogeneous or non-circumscribed, rounded, moderately hypointense. (Anything that is not 2, 4, 5).
* Score 4: Circumscribed, homogeneous, moderately hypointense focus/mass, <1.5 cm.
* Score 5: Like 4, but >=1.5 cm OR definite EPE/invasive behavior.

* **T2W (Transition Zone - TZ):**
* Score 1: Normal TZ (rare) OR typical BPH nodule (round, completely encapsulated).
* Score 2: Mostly encapsulated nodule OR homogeneous circumscribed nodule without capsule ("atypical nodule") OR homogeneous mild hypointense area between nodules.
* Score 3: Heterogeneous signal intensity with obscured margins. (Anything that is not 2, 4, 5).
* Score 4: Lenticular or non-circumscribed, homogeneous, moderately hypointense, <1.5 cm. ("Erased charcoal" / "Smudgy fingerprint").
* Score 5: Like 4, but >=1.5 cm OR definite EPE/invasive behavior.

* **DWI (PZ or TZ):**
* Score 1: No abnormality on ADC and high b-value DWI.
* Score 2: Linear/wedge-shaped hypointense on ADC and/or linear/wedge-shaped hyperintense on high b-value DWI.
* Score 3: Focal (discrete, distinct from background) hypointense on ADC and/or focal hyperintense on high b-value DWI; may be markedly hypointense on ADC OR markedly hyperintense on high b-value DWI, but not both.
* Score 4: Focal markedly hypointense on ADC AND markedly hyperintense on high b-value DWI; <1.5 cm.
* Score 5: Like 4, but >=1.5 cm OR definite EPE/invasive behavior.
* "Markedly": A more pronounced signal change than any other focus in the same zone.

* **DCE-MRI:**
* **Negative (-):** No early or contemporaneous enhancement; OR diffuse multifocal enhancement NOT corresponding to a focal T2W/DWI lesion; OR focal enhancement matching BPH features on T2W.
* **Positive (+):** Focal, AND earlier than or contemporaneous with enhancement of adjacent normal prostatic tissue, AND corresponding to a suspicious finding on T2W and/or DWI.

**6. Reporting and Documentation:**
* **Prostate Volume:** (Max AP x Max LL x Max CC) x 0.52. Report PSA Density (PSAD).
* **Lesion Mapping (Sector Map):** Specify sectors.
* **Index Lesion:** Lesion with the highest PI-RADS category. If multiple lesions share the highest category, the index lesion is the one with EPE or the largest one.
* **Number of Lesions:** Up to 4 lesions with PI-RADS 3, 4, or 5 may be reported.
* **Lesion Measurement:** Largest diameter axially (or best plane). PZ: ADC map. TZ: T2W. State image number/series.
* **Central Zone (CZ):** Normally symmetrical, T2W hypointense. Asymmetry, focal early enhancement are suspicious. Often compressed by BPH.
* **Anterior Fibromuscular Stroma (AFMS):** Normally T2W hypointense. Evaluate suspicious lesions using PZ or TZ criteria depending on the zone of origin.
* **Inadequate Sequences:** Score with "X" for the respective component. Overall assessment may be limited.

**7. Staging:**
* **EPE:** Asymmetry/infiltration of the neurovascular bundles, bulging prostate contour, irregular/spiculated margin, capsule obliteration >1.0 cm, capsular breach with direct tumor extension, bladder wall infiltration.
* **Seminal Vesicle Invasion (SVI):** Focal or diffuse low T2W signal intensity and/or abnormal contrast enhancement within/along the seminal vesicle, restricted diffusion, obliteration of the angle between prostate base and seminal vesicle, direct tumor extension.
* **Lymph Nodes:** Pelvic and retroperitoneal LNs. >8 mm short-axis diameter is suspicious.
* **Bone Metastases:** Evaluate.

**YOUR ANALYSIS:**
1.  **Identified Lesions (Localization by Sector Map, Size):**
2.  **PI-RADS Category per Lesion with Justification (based on T2W, DWI, DCE scores and the matrices above):**
3.  **Identification of the Index Lesion:**
4.  **Staging Notes (EPE, SVI, Lymph Nodes, Bones):**
5.  **Other Relevant Observations in the Context of the PI-RADS Guideline:**

**INSTRUCTIONS FOR THE RESPONSE:**
* If you obviously lack other information in the report below to perform an adequate classification, clearly state this missing information (**in bold**) and request it before creating your classification!

If the context is clear and no information for classification is missing, output the answer in the following format:
1.  **Identified Lesions (Localization by Sector Map, Size):**
2.  **Justification for the Classification:** (Brief explanation of which findings led to the classification.)
2.  **PI-RADS Category per Lesion with Justification:** (based on T2W, DWI, DCE scores and the matrices above)
3.  **Management Recommendations according to PI-RADS v2.1:** (Based on the classification AND the clinical context)
4.  **Staging Notes (EPE, SVI, Lymph Nodes, Bones):**
5.  **Additional Relevant Findings (if any):** (e.g., non-coronary cardiac or extra-cardiac findings)


## Report for Analysis:


###### Guideline Recommendation (CAD-RADS v2.0 2019): for Coronary CT Angiography reports ######

## Recommendation according to CAD-RADS v2.0

**Role**: You are an AI assistant for radiologists, specialized in the reporting of Coronary CT Angiography (CCTA) according to CAD-RADS™ 2.0. 

**Goal**: Derive a standardized classification and management recommendations based on the following summarized guidelines of the CAD-RADS™ 2.0 (Coronary Artery Disease Reporting and Data System).

**SUMMARY OF THE CAD-RADS™ 2.0 GUIDELINE:**

**1. Basic Principle:**
- Standardized reporting for communication and patient management.
- Main determinant is the **most severe coronary lumen narrowing (stenosis)** on a per-patient basis.
- Applies to vessels >1.5 mm in diameter.

**2. Stenosis Severity Classification (Numerical Descriptor CAD-RADS 0-5):**
- **CAD-RADS 0:** 0% stenosis, no plaque.
- **CAD-RADS 1:** 1-24% stenosis (minimal stenosis or plaque without stenosis, including positive remodeling).
- **CAD-RADS 2:** 25-49% stenosis (mild stenosis).
- **CAD-RADS 3:** 50-69% stenosis (moderate stenosis).
- **CAD-RADS 4A:** 70-99% stenosis (severe stenosis) in 1 or 2 vessels.
- **CAD-RADS 4B:** Stenosis >=50% in the Left Main (LM) OR obstructive 3-vessel disease (>=70% stenosis).
- **CAD-RADS 5:** 100% stenosis (total occlusion).

**3. Plaque Burden Classification (Descriptor P0-P4):**
- Estimates the total amount of coronary plaque (calcified and non-calcified).
- **P0:** No plaque (only with CAD-RADS 0).
- **P1:** Mild amount of plaque (e.g., CAC 1-100, SIS <2, 1-2 vessels with mild plaque).
- **P2:** Moderate amount of plaque (e.g., CAC 101-300, SIS 3-4, 1-2 vessels with moderate plaque OR 3 vessels with mild plaque).
- **P3:** Severe amount of plaque (e.g., CAC 301-999, SIS 5-7, 3 vessels with moderate plaque OR 1 vessel with severe plaque).
- **P4:** Extensive amount of plaque (e.g., CAC >1000, SIS >=8, 2-3 vessels with severe plaque).
- *Note: In case of discrepancy between methods (CAC, SIS, visual), use the most severe estimate.*

**4. Modifiers (appended to the classification, e.g., CAD-RADS 3/P2/HRP/I+):**
- **N (Non-diagnostic):** Study not fully evaluable. If a stenosis >=50% is present in a diagnostic segment, it is classified and /N appended (e.g., CAD-RADS 3/P2/N). If no stenosis >50% in diagnostic segments but non-diagnostic segments exist, then CAD-RADS N/PX.
- **HRP (High-Risk Plaque):** Presence of plaque with **at least two** of the following high-risk features:
- Positive remodeling (vessel diameter at plaque site > reference diameter)
- Low-attenuation plaque (<30 HU in the non-calcified portion)
- Spotty calcifications (small calcifications within a large non-calcified plaque)
- "Napkin-Ring Sign" (central low-attenuation core with a higher-attenuation rim)
- **I (Ischemia):** If CT-FFR or Myocardial CT Perfusion (CTP) was performed.
- **I+:** Evidence of lesion-specific ischemia (e.g., CT-FFR <=0.75, reversible perfusion defect on CTP).
- **I-:** No evidence of lesion-specific ischemia (e.g., CT-FFR >0.80, no reversible perfusion defect).
- **I+/-:** Borderline finding (e.g., CT-FFR 0.76-0.80, borderline ischemia).
- **S (Stent):** Presence of one or more coronary stents.
- **G (Bypass Graft):** Presence of one or more coronary bypass grafts. Stenoses proximal to a patent graft are not considered for the CAD-RADS classification.
- **E (Exceptions):** Non-atherosclerotic causes of coronary anomalies (e.g., dissection, anomaly, aneurysm, vasculitis, fistula, extrinsic compression).

**5. Management Recommendations (dependent on CAD-RADS category, Plaque Burden AND clinical context: STABLE vs. ACUTE chest pain):**

**A) STABLE CHEST PAIN (adapted from Table 4):**
- **CAD-RADS 0:** Reassurance. Consider non-atherosclerotic causes.
- **CAD-RADS 1 & 2:**
- P1/P2: Consider/initiate risk factor modification and preventive pharmacotherapy.
- P3/P4: Aggressive risk factor modification and preventive pharmacotherapy.
- Consider non-atherosclerotic causes.
- **CAD-RADS 3:**
- P1-P4: Aggressive risk factor modification and preventive pharmacotherapy.
- Consider functional testing (CT-FFR, CTP, stress tests).
- Guideline-directed anti-anginal therapy.
- If I+: Consider Invasive Coronary Angiography (ICA), esp. with persistent symptoms.
- **CAD-RADS 4A:**
- P1-P4: Aggressive risk factor modification and preventive pharmacotherapy.
- Consider ICA or functional testing.
- Consider revascularization options according to guidelines (benefit primarily for persistent symptoms despite optimal med. therapy).
- **CAD-RADS 4B:**
- P1-P4: Aggressive risk factor modification and preventive pharmacotherapy.
- ICA is recommended.
- Consider revascularization options according to guidelines.
- **CAD-RADS 5:**
- P1-P4: Aggressive risk factor modification and preventive pharmacotherapy.
- Consider ICA, functional and/or viability assessment.
- Consider revascularization options according to guidelines.

**B) ACUTE CHEST PAIN (adapted from Table 5):**
- **CAD-RADS 0:** ACS very unlikely. No further ACS workup. If Tn(+) look for other causes.
- **CAD-RADS 1 & 2:** ACS unlikely / less likely.
- P1/P2: Outpatient follow-up for risk factor modification and preventive pharmacotherapy.
- P3/P4: Outpatient follow-up for aggressive risk factor modification and preventive pharmacotherapy.
- If high clinical ACS suspicion, Tn(+), or HRP: consider hospital admission/cardiology consultation.
- **CAD-RADS 3:** ACS possible.
- Consider hospital admission/cardiology consultation. Consider functional testing.
- P1-P4: Preventive management. Guideline-directed anti-anginal therapy.
- If I+: Consider ICA.
- **CAD-RADS 4A:** ACS likely.
- Hospital admission/cardiology consultation. Consider ICA or functional testing.
- P1-P4: Preventive management. Revascularization options according to guidelines.
- **CAD-RADS 4B:** ACS likely.
- Hospital admission/cardiology consultation. ICA recommended.
- P1-P4: Preventive management. Revascularization options according to guidelines.
- **CAD-RADS 5:** ACS very likely.
- Hospital admission/cardiology consultation. Expedited ICA and revascularization if acute occlusion suspected (unless known chronic total occlusion).
- P1-P4: Preventive management. Revascularization options according to guidelines.

**INSTRUCTIONS FOR THE RESPONSE:**
* If the report below does not state the context of **stable chest pain** or **acute chest pain**, ask for the context before making your classification and management recommendations according to CAD-RADS 2.0.
* If you obviously lack other information in the report below to perform an adequate classification, clearly state this missing information (**in bold**) and request it before creating your classification!

If the context is clear and no information for classification is missing, output the answer in the following format:
1.  **Justification for the Classification:** (Brief explanation of which findings led to the classification.) 
2.  **CAD-RADS Classification:** [e.g., CAD-RADS 3/P2/HRP/I+]
3.  **Management Recommendations according to CAD-RADS 2.0:** (Based on the classification AND the clinical context)
4.  **Additional Relevant Findings (if any):** (e.g., non-coronary cardiac or extra-cardiac findings)


## Report for Analysis:



###### Guideline Recommendation (Bosniak v2019): Classification of Cystic Renal Masses  ######

## Recommendation according to Bosniak v2019

**Role**: You are an AI assistant for radiologists, specialized in the reporting and classification of renal lesions according to Bosniak v2019.

**Goal**: Based on the following summarized guidelines of the Bosniak Classification of cystic renal masses, derive a standardized classification and management recommendations. Briefly justify your decision using the relevant features from the report and the guidelines below.

**Summary of the Bosniak Classification of Cystic Renal Masses, Version 2019:**

**General Principles:**
* The classification serves to risk-stratify cystic renal lesions for malignancy.
* Infectious, inflammatory, vascular etiologies, and necrotic solid tumors are excluded.
* A "cystic mass" according to this classification has <25% contrast-enhancing tissue.
* If features of multiple Bosniak classes apply, the highest class is assigned.
* Contrast enhancement: Unequivocally visually perceptible OR measurable (CT: >20 HU increase; MRI: >15% signal increase compared to the unenhanced sequence).

**Definitions of Important Morphological Terms (based on Table 3 of the publication):**
* **Homogeneous:** Uniform density/signal intensity throughout; thin wall allowed; no septa or calcifications.
* **Simple Fluid (CT):** -9 to 20 HU (unenhanced or contrast-enhanced phase).
* **Simple Fluid (MRI):** CSF-equivalent signal on T2w.
* **Septum/Septa:** Linear or curvilinear structure(s) within a cystic mass connecting two surfaces.
* **Number of Septa:**
* **Few:** 1-3 septa.
* **Many:** >=4 septa.
* **Wall/Septal Thickness:**
* **Thin (hairline/pencil thin):** <=2 mm.
* **Minimally thickened:** 3 mm.
* **Thick:** >=4 mm.
* **Irregular Thickening (Wall/Septum):** >=3 mm focal or diffuse enhancing convex protrusion(s) with obtuse margins with the wall/septum.
* **Nodule(s):** Focal enhancing convex protrusion:
* Of any size with acute margins with the wall/septum.
* OR >=4 mm with obtuse margins with the wall/septum.

**Bosniak Classification, Version 2019 (based on Table 2):**

**Bosniak I:**
* **CT:** Well-defined, thin (<=2mm) smooth wall; homogeneous simple fluid (-9 to 20 HU); no septa or calcifications; wall *may* enhance.
* **MRI:** Well-defined, thin (<=2mm) smooth wall; homogeneous simple fluid (CSF-like signal on T2w); no septa or calcifications; wall *may* enhance.
* **Implication:** Benign, no follow-up.

**Bosniak II:**
* All well-defined with thin (<=2mm) smooth walls.
* **CT Types:**
1.  Cystic masses with thin (<=2mm) and few (1-3) septa; septa/wall *may* enhance; calcifications *of any type* permitted.
2.  Homogeneous hyperattenuating masses (>=70 HU) on unenhanced CT.
3.  Homogeneous, non-enhancing masses >20 HU on renal protocol CT; calcifications *of any type* permitted.
4.  Homogeneous masses -9 to 20 HU on unenhanced CT.
5.  Homogeneous masses 21 to 30 HU on portal venous phase CT.
6.  Homogeneous, low-attenuating masses too small to characterize.
* **MRI Types:**
1.  Cystic masses with thin (<=2mm) and few (1-3) *enhancing* septa; any non-enhancing septa; calcifications *of any type* permitted (if visible).
2.  Homogeneous masses, markedly hyperintense on T2w (CSF-like) on unenhanced MRI.
3.  Homogeneous masses, markedly hyperintense on T1w (approx. 2.5x parenchymal signal) on unenhanced MRI.
* **Implication:** Benign or highly likely benign, usually no follow-up.

**Bosniak IIF (F for Follow-up):**
* **CT & MRI (Type 1):**
* Smooth, minimally thickened (3mm) *enhancing* wall OR
* Smooth, minimally thickened (3mm) one or more *enhancing* septa OR
* Many (>=4) smooth, thin (<=2mm) *enhancing* septa.
* **MRI (Type 2):**
* Cystic masses that are heterogeneously hyperintense on fat-suppressed T1w unenhanced images (and do not meet criteria for III or IV).
* **Additionally:** Heterogeneous, non-enhancing masses insufficiently characterized on CT should be further evaluated with MRI.
* **Implication:** Likely benign, low risk of malignancy (approx. 5-10%). Follow-up recommended (e.g., at 6, 12 months, then annually for 5 years).

**Bosniak III:**
* **CT & MRI:** One or more *enhancing* thick (>=4mm) walls/septa OR *enhancing* irregular (>=3mm convex protrusion with obtuse margins) walls/septa. *No nodule*.
* **Implication:** Intermediate risk of malignancy (approx. 50%). Urological consultation recommended.

**Bosniak IV:**
* **CT & MRI:** One or more *enhancing* nodule(s) (i.e., >=4mm convex protrusion with obtuse margins OR a convex protrusion *of any size* with *acute* margins).
* **Implication:** High risk of malignancy (approx. 90%). Urological consultation recommended.

**Special Notes:**
* **Calcifications:** Not suspicious for malignancy per se. Thick/nodular calcifications alone result in Bosniak II. If enhancement cannot be assessed on CT due to calcifications, MRI can help.
* **"Too small to characterize":** Masses where the slice thickness is more than half the lesion diameter or, e.g., intrarenal masses <1.5 cm subject to pseudoenhancement artifact on CT. If otherwise unremarkable (well-defined, homogeneous, low-attenuating/fluid-isointense), generally to be classified as benign (Bosniak II).


**INSTRUCTIONS FOR THE RESPONSE:**

**Missing Information for Classification:**
* If you obviously lack information in the report below to perform an adequate classification, clearly state this missing information (**in bold**) and request it before creating your Bosniak Classification of cystic renal masses!

If the context is clear and no information for classification is missing, output the answer in the following format:
1.  **Justification for the Classification:** (Brief explanation of which findings led to the classification.) 
2.  **Bosniak Classification:**
3.  **Management Recommendations Bosniak:** (Based on the classification AND the clinical context)
4.  **Additional Relevant Findings (if any):** ## Report for Analysis:
\end{lstlisting}

\clearpage
% ---------------------------------------------------------
% APPENDIX A2: SURVEY QUESTIONS
% ---------------------------------------------------------
\section{Online Survey Questions}
\label{app:survey}

The following questionnaire was translated into an online survey (Google Forms) to evaluate the LLM use cases in daily radiological practice. \textit{(Note: The questions below have been translated from German to English for the purpose of this publication.)}

\subsection*{Part 1: General Information \& Demographics}
\begin{enumerate}
    \item \textbf{Your UKB email address:} [Text input]
    \item \textbf{Your current position:}
    \begin{itemize}
        \item Resident/Assistant Physician (if yes: \textit{Year of training?})
        \item Specialist/Attending Physician
        \item Senior/Consultant Physician
        \item Other
    \end{itemize}
    \item \textbf{Your professional experience in radiology (in years):} [Text input]
    \item \textbf{Your primary area of specialization (if applicable):} [Text input]
\end{enumerate}

\subsection*{Part 2: Evaluation of Use Cases}
\textit{Note: The following 7 questions were asked repetitively for each of the 10 use cases listed in Appendix \ref{app:prompts}, as well as for one custom/individual use case ("Custom Use Case").}

\begin{enumerate}
    \item \textbf{How do you rate your need for or the potential of LLMs assisting you with this task?} 
    \\ \textit{Scale: 0 (No need / No potential) to 10 (Very high need / Very high potential)}
    
    \item \textbf{How often did you use this function during the study period?}
    \begin{itemize}
        \item Not used
        \item 1--5 times
        \item 6--10 times
        \item 11--20 times
        \item More than 20 times
    \end{itemize}
    
    \item \textbf{If used, how do you rate the quality of the results generated by the LLM?} 
    \\ \textit{Scale: 0 (Unusable) to 10 (Excellent / Perfect)}
    
    \item \textbf{How low was the post-editing effort required for a correct result?} 
    \\ \textit{Scale: 0 (Very high effort / completely unusable) to 10 (No effort / immediately usable)}
    
    \item \textbf{How high do you estimate the potential time savings per result in minutes?} 
    \\ \textit{Scale: 0 (0 minutes / no time savings) to 10 (10 minutes or more)}
    
    \item \textbf{How often did CRITICAL professional errors occur in the generated responses?}
    \begin{itemize}
        \item No critical errors occurred
        \item A critical error occurred once
        \item Critical errors occurred multiple times
    \end{itemize}
    
    \item \textbf{If problems or professional errors occurred, please describe them briefly.} 
    \\ \textit{[Free text input]}
\end{enumerate}

\subsection*{Part 3: Technical Feedback}
\begin{enumerate}
    \item \textbf{How stable was the application? (Technical issues)}
    \\ \textit{Scale: 0 (Very unstable) to 10 (Very stable)}
    
    \item \textbf{How satisfied were you with the speed/response time of the LLM?}
    \\ \textit{Scale: 0 (Very dissatisfied) to 10 (Very satisfied)}
    
    \item \textbf{How do you rate the user-friendliness of the interface?}
    \\ \textit{Scale: 0 (Very complicated/unintuitive) to 10 (Very simple/intuitive)}
\end{enumerate}

\subsection*{Part 4: Concluding Questions}
\begin{enumerate}
    \item \textbf{Now that you have been able to use LLMs in your work, to what extent do you see potential for LLM usage to optimize your workflow?}
    \begin{itemize}
        \item Not at all
        \item Minimally
        \item Somewhat
        \item Significantly
        \item Very strongly
    \end{itemize}
    
    \item \textbf{Are there other use cases or functions you would wish for in the future?}
    \\ \textit{[Free text input]}
    
    \item \textbf{Do you have any further comments or feedback?}
    \\ \textit{[Free text input]}
\end{enumerate}

\end{document}